%% file: main.tex
\documentclass[12pt]{iopart}

\usepackage{iopams}  

\expandafter\let\csname equation*\endcsname\relax
\expandafter\let\csname endequation*\endcsname\relax
\usepackage{amsmath}
\usepackage{hyperref}

\usepackage{comment}
\usepackage{bbm}
\usepackage{tensor}
\usepackage{mathtools}
\usepackage{xcolor}						
\usepackage{etoolbox}
\usepackage{enumitem}
\usepackage{accents}

\usepackage{etoolbox}
\makeatletter
\newcommand{\mainmatter}{%
  \setcounter{footnote}{0}%
  \patchcmd{\@makefntext}{\fnsymbol}{\arabic}{}{}%
  \patchcmd{\@thefnmark}{\fnsymbol}{\arabic}{}{}%
  \def\@makefnmark{\textsuperscript{\arabic{footnote}}}%
}
\makeatother

\input{lib.tex}

\usepackage{amsthm}
\newtheorem{theorem}{Theorem}[section]

\newtheorem{definition}[theorem]{Definition}

\newtheorem*{realityI}{Reality condition I}
\newtheorem*{realityII}{Reality condition II}

\begin{document}

\title[Revisiting LQG with selfdual variables: Hilbert space and first reality condition]{Revisiting loop quantum gravity with selfdual variables: Hilbert space and first reality condition}

\author{Hanno Sahlmann$^1$ and Robert Seeger$^2$}

\address{Friedrich-Alexander-Universit\"at Erlangen-N\"urnberg (FAU)\\ Institute for Quantum Gravity, Staudtstra{\ss}e 7/B2, 91058 Erlangen, Germany}
\ead{$^1$ hanno.sahlmann@gravity.fau.de, $^2$ robert.seeger@gravity.fau.de}
\vspace{10pt}
\begin{indented}
\item[]October 2023
\end{indented}

\begin{abstract}
We consider the quantization of gravity as an $\SLtwoC$ gauge theory in terms of Ashtekar's selfdual variables and reality conditions for the spatial metric (RCI) and its evolution (RCII). We start from a holomorphic phase space formulation. It is then natural to push for a quantization in terms of holomorphic wave functions. Thus we consider holomorphic cylindrical wave functions over $\SLtwoC$ connections.  

We use an overall phase ambiguity of the complex selfdual action to obtain Poisson brackets that mirror those of the real theory. We then show that there is a representation of the corresponding canonical commutation relations the space of holomorphic cylindrical functions. We describe a class of cylindrically consistent measures that implements RCI. We show that spin networks with $\SUtwo$ intertwiners form a basis for gauge invariant states. They are still mutually orthogonal, but the normalisation is different than for the Ashtekar-Lewandowski measure for $\SUtwo$.   

We do not consider RCII in the present article. Work on RCII is ongoing and will be presented elsewhere. 
\end{abstract}

%\maketitle
\mainmatter
\section{Introduction}\label{sec:introduction}

Ashtekar's selfdual variables \cite{sen_gravity_1982,ashtekar_new_1986,ashtekar_new_1987,ashtekar_gravitational_2021} are interesting for many reasons. For one thing, the standard model of particle physics is chiral, and so a chiral description of gravity seems natural. They also lead to a simple canonical formulation, with canonical variables $A$ (an $\SLtwoC$ connection) and $E$ (a densitized triad)  that are pull-backs from spacetime, and a first order system of constraints, with the  Hamiltonian constraint taking a particularly simple form. Therefore, they have been considered a promising starting point for canonical quantum gravity. The big technical problem with this application is that, in Lorentzian signature, the variables are complex. Reality conditions, in part non-polynomial, ensure they describe a real spacetime metric. 

Therefore, most progress in loop quantum gravity (see for example \cite{thiemann_modern_2007,ashtekar_background_2004,rovelli_spin_2013}) has been made using a formulation in terms of a real valued $\SUtwo$ connection, the Ashtekar-Barbero connection \cite{barbero_g_real_1995,immirzi_real_1997}. However, there are hints from the $\SUtwo$ theory that a more fundamental quantum theory with $\SLtwoC$ gauge symmetry might exist, see \cite{frodden_black-hole_2014,jibril_analytic_2015,eder_towards_2022,eder_chiral_2023} in the context of black holes and \cite{achour_loop_2015,ben_achour_covariance_2018} in the context of cosmology. 

Previous work to quantize gravity in terms of selfdual variables took various routes. Early on, it was suggested to use holomorphic wave functions \cite{ashtekar_self_1992,ashtekar_recent_1992,tate_algebraic_1993} and the Segal-Bargmann transform \cite{ashtekar_coherent_1996}. 
A very interesting and somewhat related suggestion is to use a kind of Wick transform \cite{thiemann_account_1995,thiemann_reality_1996} that relates the constraints in the selfdual theory to those for the real variables. A crucial ingredient here is the construction of a complexifier, an operator that generates the transformation. Recently, some remarkable new ideas and progress in this direction have been reported \cite{varadarajan_euclidean_2019}. 

Another important proposal to make use of the selfdual variables for quantization is \cite{wieland_complex_2012}, see also \cite{wieland_twistorial_2012}. The idea is to write the real theory in terms of a set of (partially second class) constraints on a kinematical phase space for a $\SLtwoC$ connection and its canonical conjugate. The connection can be thought of as the selfdual part of a complex connection. Intriguingly, one of the constraints, a kind of reality condition for the metric, takes the form of the linear simplicity constraints of spinfoam gravity. 

It has also been demonstrated that selfdual variables can be used to quantize cosmological sectors of general relativity, see  \cite{wilson-ewing_loop_2015,achour_loop_2015,ben_achour_covariance_2018,eder_supersymmetric_2021}. Furthermore, two very recent and exciting developments are a consistent simplicial regularization of the constraints of the selfdual theory \cite{wieland_simplicial_2023} and an inner product which solves the reality conditions for selfdual variables, based on a formal path integral \cite{alexander_inner_2022}.

With the present work, we would like to add to the conversation around quantum gravity using selfdual variables. We will go back to the old suggestions to use analytic methods for the quantization \cite{ashtekar_self_1992,ashtekar_recent_1992,tate_algebraic_1993,thiemann_reality_1996,ashtekar_coherent_1996}, in that we will use analytic wave-functions of the selfdual connection $A$ as quantum states. This is suggested by the nature of the phase space derived from the selfdual action in Minkowski signature, as a space of holomorphic functions of $A$ and its canonically conjugate.\footnote{The pre-symplectic structure is degenerate for real and imaginary parts of $A$ treated as separate fields, see for example \cite{sahlmann_revisiting_classical_theory}. In this situation, one can either modify the symplectic structure, as effectively done in \cite{wieland_complex_2012}, or one can interpret it as a phase space of holomorphic functions, a viewpoint we will adopt here.} As a consequence, we can not treat the reality conditions as constraints, as they would not be holomorphic. Rather, we view them as conditions on the representation of the canonical commutation relations, for example in the form of adjointness relations. 

In \cite{sahlmann_revisiting_classical_theory}, we have made the observation, that due to the complex nature of the selfdual action, there is an arbitrary overall phase $\lambda$. While it has essentially no effect in the classical theory, its effect on the quantum theory is rather profound, as it shows up in the canonical commutation relations. For example, we will see that for the choice $\lambda=i$ we will find a class of measures on the space of holomorphic cylindrical functions on $\SLtwoC$-connections that obeys the reality condition on the metric (RCII), but it is clear that our methods will not lead to such measures for the case $\lambda=1$. The latter form for the commutation relations is the standard form found in the literature, but there are notable exceptions: \cite{thiemann_reality_1996} uses Poisson brackets for the complex variables that would correspond our $\lambda=i$, and also the Poisson brackets of \cite{wieland_complex_2012} (for the fields, not their conjugates) would look like ours for $\lambda=i$. 

Using holomorphic wave functions, along with setting $\lambda=i$ and some other choices we make, has the interesting effect of bringing the quantum theory we devise very close to that of the real theory. We will encounter consistent families of measures on cylindrical functions, and spin networks labeled with $\SUtwo$ irreducible representations. But the holonomies we consider are still $\SLtwoC$-valued.   

In the present work, we will treat in detail only RCI. Work is in progress regarding RCII and a publication will be forthcoming \cite{sahlmann_revisiting_RCII}. 

The structure of the article is as follows: In section \ref{sec:classicaltheory}, we review the classical formulation that we are starting from, and in particular the $\lambda$-ambiguity and the reality conditions. In section \ref{sec:analytic_functions}, we will discuss the mathematical structures needed for the quantum theory, such as analytic functions on $\SLtwoC$, the holomorphic invariant derivatives, and analytic representations. 
Section \ref{sec:hfalgebra} sketches the definition of the holonomy-flux algebra in the $\SLtwoC$ case. Section \ref{sec:main} contains the bulk of this work: In \ref{sec:adjointness}, RCI is translated into an adjointness relation for the right- and left-invariant vector fields, and in \ref{sec:realitycondition} into a condition on the Radon-Nikodym derivative of the measure on one edge relative to the Haar measure on $\SLtwoC$. In section \ref{sec:solution}, a class of Radon-Nikodym derivatives is found that solve the reality conditions. Sections \ref{sec:holomorphic_irreps} -- \ref{sec:spin_networks_gauge_invariance} then establish various properties for the measure, in particular cylindrical consistency, so that a measure on generalized $\SLtwoC$-connections is obtained.  
We close with a discussion of the results and an outlook on further work in section \ref{sec:discussion}.

We use the metric signature $(-,+,+,+)$.

\section{Classical theory}\label{sec:classicaltheory}

In this section we want to lay out the classical theory of complex general relativity, which is described in detail in \cite{sahlmann_revisiting_classical_theory}, and whose quantisation is the subject of this work. 

From the the selfdual part of the Palatini action based on the complex Lorentz group $\LGC$, we obtain the following action for the $\sltwoC$-valued vector-density $\tE^a_i$, the electric field, and $A^i_a$, the Ashtekar's $\SLtwoC$ connection: 
\begin{equation}
    S=\int_\RR\grad{t}\int_M \grad{^3x} \rb{\frac{2\lambda}{\kappa i} \tE^a_i\mathop{\curlyL_t}(A^i_a)-\rb{NC+N^aC_a+(A\cdot t)^i G_i}}.
\end{equation}
The Hamilton, diffeomorphism and Gau{\ss} constraints have the usual form 
\begin{equation}\label{eq:classical_constraints} 
\begin{aligned}
    N C &= -N\frac{\lambda s_e}{\sqrt{q} \kappa s_E}\lc{_{k}^{ij}}\tE^a_i\tE^b_j F^k_{ab},\\
    N^a C_a &= \frac{2\lambda}{i\kappa} N^a  \tE^b_i F^{i}_{ab},\\
    (A\cdot t)^i G_i &= - (A\cdot t)^i \frac{2\lambda }{i\kappa} \mathop{D_a}(\tE^a_i).
\end{aligned}    
\end{equation}
$s_e$ and $s_E$ are independent signs related to the orientation of spacetime tetrads and spatial triads, respectively. They do not play a role in the quantum theory, which is described in this work. For a detailed description of their origin, see \cite{sahlmann_revisiting_classical_theory}. 

Relevant for the present work, however, is the complex factor $\lambda$. Important in the following are the specific values $\lambda=1$ and $\lambda=i$. As described in \cite{sahlmann_revisiting_classical_theory}, it does not play a significant role for the classical theory. Remarkably, it is completely irrelevant for the reality conditions and for $\lambda=1$ only causes a slightly unconventional interpretation of the classical theory, once reality is restored classically. 
For the quantisation in the present work, however, the value of $\lambda$ plays a crucial role. 

The Poisson relation for Ashtekar's variables are given by 
\begin{equation}\label{eq:Poisson_Relation_AE}
    \PB{A^i_a(x)}{\tE^b_j(y)}=\frac{i\kappa}{2\lambda}\delta^b_a\delta^i_j\delta(x,y).
\end{equation}
Therefore, although $\lambda$ does not affect the dynamics of the classical theory, as Poisson relation and constraints have an inverse dependence, it affects the Poisson relation and therefore the canonical commutation relation in the quantum theory. 
Specifically, $\lambda=1$ corresponds to the standard Poisson relation of selfdual gravity throughout the literature, e.\,g.\cite{ashtekar_new_1986,baez_gauge_1994,ashtekar_lectures_1998,ashtekar_gravitational_2021}, characterised by the factor of $i$. In contrast, $\lambda=i$ removes this feature of selfduality and enforces a Poisson relation that is similar to real Ashtekar-Barbero gravity, even though the connection and electric field are complex. Similar Poisson relations, although in different context and with different origin, are used in \cite{thiemann_account_1995,thiemann_reality_1996} and \cite{wieland_complex_2012}.\footnote{In \cite{thiemann_account_1995,thiemann_reality_1996} they arise from a canonical transformation, which maps real Ashtekar-Barbero variables to a specific form of complex Ashtekar variables. In \cite{wieland_complex_2012} they arise from expressing the real Palatini-Holst action in terms of a complex action and its complex conjugate. However, as additional Poisson relations are taken into account, no holomorphic point of view is taken.}  

We want to make clear that all of the above has to be understood as a holomorphic description of complex gravity and that there are no Poisson relations for real and imaginary parts of $A^i_a$ and $\tE^b_j$. Again, see \cite{sahlmann_revisiting_classical_theory} for details.

The reality of the classical theory is retained by imposing reality conditions on the spatial metric and its time evolution.
As the electric fields serve as densitised triads of the theory, the spatial metric $q_{ab}$ of complex gravity is obtained via contraction with the internal Euclidean metric:
\begin{equation}\label{eq:densitised_metric}
    q q^{ab}=\tE^a_i \tE^b_j \delta^{ij}.
\end{equation}
In this way, demanding reality of the spatial metric is transferred to the electric field.

In order to keep the spatial metric real under time evolution, the Ashtekar connection becomes relevant. Here it is important that it can be decomposed into 
\begin{equation}
    A^j_a=\Gamma^j_a+iK^j_a,
\end{equation}
where $\Gamma^j_a$ is the spin connection compatible with the densitised triads, while $K^j_a$ is the complex version of the extrinsic curvature.   

The reality conditions are:
\begin{realityI}
The following statements are equivalent:\begin{enumerate}[label=(\roman*)]
    \item \label{thm:E_real} $\tE^a_i$ is real, 
    \item \label{thm:qq_posdef} $qq^{ab}=\delta^{ij}\tE^a_i\tE^b_j$ is real and positive definite, 
    \item \label{thm:q_real} $q^{ab}$ is real.
\end{enumerate}
\end{realityI}

\begin{realityII}
Let the first reality condition hold. Then the following statements are equivalent:
\begin{enumerate}[label=(\roman*)]
\item \label{thm:K_real} $K^i_a$ is real,
\item \label{thm:qqdot_real} $(qq^{ab})^{\boldsymbol{\cdot}}=(\delta^{ij}\tE^a_i\tE^b_j)^{\boldsymbol{\cdot}}$ is real,
\item \label{thm:qdot_real} $\dot{q}^{ab}$ is real.
\end{enumerate}
\end{realityII}

Note that we omitted reality up to gauge rotations of $\tE^a_i$ and $K^b_j$, compared to the statements in \cite{sahlmann_revisiting_classical_theory}. This simplifies the implementation of reality conditions in the quantum theory, as 
we have definite conditions for the quantum counterparts of electric field and extrinsic curvature. Nevertheless it takes nothing away from the fact that we want to impose quantum reality conditions on a quantum theory based on complex variables.

\section{Analytic functions, invariant vector fields and holomorphic irreducible representations of $\SLtwoC$}\label{sec:analytic_functions}

The determining character of the classical theory, as described above, is holomorphicity with respect to Ashtekar's variables. For the quantum theory we therefore want to keep this and perform the quantisation in a holomorphic fashion.  

As we will show later in this work, the quantum theory will be constructed using a Hilbert space of analytic functions on $\SLtwoC$, on which we will act with holomorphic invariant vector fields. All of this will make use of holomorphic irreducible representations of $\SLtwoC$. 

The purpose of the following sections is to introduce these notions, before we get to the actual quantisation.

\subsection{Analytic functions of $\SLtwoC$}
We start with the notion of analytic functions of $\SLtwoC$. We will treat them simply as complex analytic functions of four variables, being the four matrix elements $\T{g}{^I_J}$ with $I,J=1,2$ of $g\in\SLtwoC$. 
In the following we will however use the general form
\begin{equation}\label{eq:g_abcd}
    g=\begin{pmatrix}
a & b \\
c & d 
\end{pmatrix}
\end{equation}
with $a,b,c,d\in\CC$ and $\det(g)=ad-bc=1$. Therefore the holomorphic and antiholomorphic derivatives with respect to the matrix elements are given by
\begin{equation}
\begin{aligned}
        &\frac{\partial}{\partial x}=\frac{1}{2}\rb{\frac{\partial}{\partial \mathrm{Re}(x)}-i\frac{\partial}{\partial \mathrm{Im}(x)}},\\
    &\frac{\partial}{\partial \bar{x}}=\frac{1}{2}\rb{\frac{\partial}{\partial \mathrm{Re}(x)}+i\frac{\partial}{\partial \mathrm{Im}(x)}},
\end{aligned}
\end{equation}  
with $x=a,b,c,d$, and 
\begin{equation}
    \frac{\partial \bar{x}}{\partial x}=0=\frac{\partial x}{\partial \bar{x}}.
\end{equation}
Note that we will use the terms analytic and holomorphic functions interchangeably. \footnote{Holomorphic functions of several variables are by definition complex analytic. They are  also holomorphic in all variables separately. Vice versa, a function holomorphic in each variable separately is analytic and hence holomorphic in all variables.}   

\begin{definition}[Analytic functions of $\SLtwoC$]\label{def:analytic_functions}
\begin{enumerate}
    \item 
    The function $\Psi:\SLtwoC\rightarrow\CC$ is called holomorphic in $g\in\SLtwoC$  if and only if $\forall x\in\{a,b,c,d\}$
    \begin{equation}
        \frac{\partial}{\partial \bar{x}}\Psi(g)=0.
    \end{equation}
    \item 
    The function $\Psi:\SLtwoC\rightarrow\CC$ is called anti-holomorphic in $\bar{g}\in\SLtwoC$  if and only if $\forall x\in\{a,b,c,d\}$
    \begin{equation}
        \frac{\partial}{\partial x}\Psi(g)=0.
    \end{equation}
\end{enumerate}    
\end{definition}
In order to fix the notation, we will denote the holomorphic functions of $\SLtwoC$ as $\mathfrak{H}(\SLtwoC)$ and the anti-holomorphic functions as $\overline{\mathfrak{H}}(\SLtwoC)$.
It is straightforward to extend this definition to functions of more than one element of $\SLtwoC$, by considering the corresponding (anti-)holomorphic derivatives in the respective matrix elements.

\subsection{Holomorphic invariant vector fields} 
With its generators for boost and rotations, $\SLtwoC$ gives rise to six left- and right-invariant vector fields. However considering $\sltwoC$ as the complexification of $\sutwo$, i.\,e., the complex span of the standard generators, we can introduce three holomorphic left- and right-invariant vector fields, respectively, and in addition their anti-holomorphic counterparts. 

In the following definition we use the sign convention of \cite{ashtekar_background_2004} for invariant vector fields.

\begin{definition}[Holomorphic invariant vector fields]\label{def:hol_vf}
    \begin{enumerate}
        \item Let $\Psi(g)$ be a holomorphic function of $\SLtwoC$, $\tau_j=\sigma_j/(2i)$ $j=1,2,3$ the generators of $\sltwoC$ and $\varepsilon\in\CC$. Then
        \begin{equation}\label{eq:hol_left_vf}
            (L_i\Psi)(g)=\frac{d}{d\varepsilon}\Big\vert_{\varepsilon=0}\Psi(g e^{\varepsilon\tau_i})
        \end{equation}
        and
        \begin{equation}\label{eq:hol_right_vf}
            (R_i\Psi)(g)=\frac{d}{d\varepsilon}\Big\vert_{\varepsilon=0}\Psi(e^{-\varepsilon\tau_i}g)
        \end{equation}
        are called holomorphic left- and right-invariant vector fields of $\SLtwoC$.
        \item Let $\Psi(\bar{g})$ be a anti-holomorphic function of $\SLtwoC$, $\tau_j$ and $\varepsilon$ as above. Then
        \begin{equation}
            (\bar{L}_i\Psi)(\bar{g})=\frac{d}{d\bar{\varepsilon}}\Big\vert_{\bar{\varepsilon}=0}\Psi(\overline{ge^{\varepsilon\tau_i}} )
        \end{equation}
        and
        \begin{equation}
            (\bar{R}_i\Psi)(\bar{g})=\frac{d}{d\bar{\varepsilon}}\Big\vert_{\bar{\varepsilon}=0}\Psi(\overline{e^{-\varepsilon\tau_i}g})
        \end{equation}
        are called anti-holomorphic left- and right-invariant vector fields of $\SLtwoC$.
    \end{enumerate}
\end{definition}

Note that (anti-)holomorphicity not only refers to the derivatives in the parameter $\varepsilon$, but also to the action of the vector fields. With the definition above we have $\bar{L}_i\Psi(g)=0$ as well as $L_i\Tilde{\Psi}(\bar{g})=0$, and likewise for the action of $R_i$. 

This distinction now plays an important role when we want to parametrise non-holomorphic functions in terms of $g$ and $\bar{g}$, which are treated independently. On such functions, we can recover invariant vector fields for boosts and rotations. For this we consider $\varepsilon=\alpha+i\beta$ with $\alpha,\beta\in\RR$. So $\alpha$ corresponds to rotations, while $i\beta$ corresponds to boosts.
The (anti-)holomorphic derivatives decompose according to 
\begin{equation}
        \frac{d}{d\varepsilon}=\frac{1}{2}\rb{\frac{\partial}{\partial\alpha}-i\frac{\partial}{\partial\beta}},\quad
        \frac{d}{d\bar{\varepsilon}}=\frac{1}{2}\rb{\frac{\partial}{\partial\alpha}+i\frac{\partial}{\partial\beta}}.
\end{equation}
This is of course equivalent to 
\begin{equation}\label{eq:rotational_and_boost_derivatives}
        \frac{\partial}{\partial\alpha}=\frac{d}{d\varepsilon}+\frac{d}{d\bar{\varepsilon}},\quad
        \frac{\partial}{\partial\beta}=\frac{1}{i}\rb{\frac{d}{d\varepsilon}-\frac{d}{d\bar{\varepsilon}}}.
\end{equation}

Now acting on functions $\Psi(g,\bar{g})$ we see for the left invariant vector fields that 
\begin{equation}
\begin{aligned}
    \rb{L_i+\bar{L}_i}\Psi(g,\bar{g})&=\frac{d}{d\varepsilon}\Big\vert_{\varepsilon=0}\Psi(ge^{\epsilon\tau_i},\bar{g})+\frac{d}{d\bar{\varepsilon}}\Big\vert_{\bar{\varepsilon}=0}\Psi(g,\overline{ge^{\epsilon\tau_i}})\\
    &=\rb{\frac{d}{d\varepsilon}+\frac{d}{d\bar{\varepsilon}}}\Big\vert_{\bar{\varepsilon}=0}\Psi(ge^{\epsilon\tau_i},\overline{ge^{\epsilon\tau_i}})\\
    &=\frac{\partial}{\partial\alpha}\Big\vert_{\alpha=0}\Psi(ge^{\alpha\tau_i},\overline{ge^{\alpha\tau_i}}),
\end{aligned}
\end{equation}
where from the first to the second line we used that the added shift are not seen by the respective (anti-)holomorphic derivative. Similar calculations for subtracting $\bar{L}_i$ from $L_i$ and the right-invariant counterparts are obvious. This allows to introduce the rotational and boost invariant vector fields, expressed by the holomorphic ones:
\begin{equation}\label{eq:rotational_and_boost_invariant_vector_fields}
    \begin{aligned}
        L_i^\mathrm{R}\Psi(g,\bar{g})&= \frac{d}{d\alpha}\Big\vert_{\alpha=0}\Psi(ge^{\alpha\tau_i},\overline{ge^{\alpha\tau_i}})=\rb{L_i+\bar{L}_i}\Psi(g,\bar{g}),\\
        R_i^\mathrm{R}\Psi(g,\bar{g})&= \frac{d}{d\alpha}\Big\vert_{\alpha=0}\Psi(e^{-\alpha\tau_i}g,\overline{e^{-\alpha\tau_i}g})=\rb{R_i+\bar{R}_i}\Psi(g,\bar{g}),\\
        L_i^\mathrm{B}\Psi(g,\bar{g})&= \frac{d}{d\beta}\Big\vert_{\beta=0}\Psi(ge^{i\beta\tau_i},\overline{ge^{i\beta\tau_i}})=\frac{1}{i}\rb{L_i-\bar{L}_i}\Psi(g,\bar{g}),\\
        R_i^\mathrm{B}\Psi(g,\bar{g})&= \frac{d}{d\beta}\Big\vert_{\beta=0}\Psi(e^{-i\beta\tau_i}g,\overline{e^{-i\beta\tau_i}g})=\frac{1}{i}\rb{R_i-\bar{R}_i}\Psi(g,\bar{g}).
    \end{aligned}
\end{equation}

Finally we want to mention that the holomorphic invariant vector fields introduced here, as a representation of $\sltwoC$, have to be understood as the complexified representation of $\sutwo$, given by its invariant vector fields.

\subsection{Holomorphic representations of $\SLtwoC$}\label{sec:holomorphic_irreps}

Real LQG heavily relies on the irreducible representations of $\SUtwo$. With the use of the Peter-Weyl theorem, the spin network functions, that form an orthonormal basis of the Ashtekar-Lewandowski Hilbert space, in itself are formed from the irreducible representations, as basic building blocks, connected via intertwiners, therefore ensuring gauge invariance under the $\SUtwo$ Gau{\ss} constraint. So to a large extend, the irreps of $\SLtwoC$ characterise the quantum theory, based on real Ashtekar-Barbero variables. 

Working, like here, with $\SLtwoC$ instead of $\SUtwo$, we have to deal with the more complicated representation theory of the non-compact group. Broad classes of irreducible representations are the infinite dimensional and the finite dimensional ones. All unitary irreducible representations are among the infinite dimensional ones, due to $\SLtwoC$ being non-compact. 
The finite dimensional irreps of $\SLtwoC$ are labelled by two spins $(j_1,j_2)$. 
As we will show now, neither of this classes of representations are holomorphic, in general.  
The necessary representation theory is compiled and presented in \cite{martin-dussaud_primer_2019}. We refer to this work and the references therein. On the Hilbert space $\curlyL^2(\CC)$ with an appropriate measure, we consider
\begin{equation}\label{eq:SLtwoC_realisations}
    g\cdot f(z)=(g_{12}z+g_{22})^{\frac{m}{2}-\frac{i\rho}{2}-1}(\overline{g_{12}z+g_{22}})^{-\frac{m}{2}+\frac{i\rho}{2}-1}f\rb{\frac{g_{11}z+g_{21}}{g_{12}z+g_{22}}}.
\end{equation}
For $m\in\mathbb{Z}$, $\rho\in\CC$ and $\Im(\rho)\geq0$ and $\rho^2\neq-(|m|+2n)^2$ with $n\in\mathbb{N_+}$, this is a realisation of the infinite dimensional representations. However for $\rho^2=-(|m|+2n)^2$ and $n\in\mathbb{N_+}$ and restricted to polynomials of degree $p=\frac{m}{2}-\frac{i\rho}{2}-1$ and $q=-\frac{m}{2}+\frac{i\rho}{2}-1$ in $z$ and $\bar{z}$, respectively, it is a realisation of the finite dimensional representation $(\frac{p}{2},\frac{q}{2})$. From \eref{eq:SLtwoC_realisations}, as it contains the complex conjugate, we readily see how neither the infinite nor the finite dimensional representations are in general holomorphic. 

From \eref{eq:SLtwoC_realisations} it is clear that the representation only has a chance to be holomorphic or anti-holomorphic, if $q=0$ or $p=0$, respectively. Indeed these representations turn out to be holomorphic, respectively, anti-holomorphic as one can see in the following way. 
Weyl's unitary trick tells us that the irreducible representations of $\SUtwo$ are in fact holomorphic irreducible representations of $\SLtwoC$ as well \cite{knapp_representation_2001,martin-dussaud_primer_2019}. They are however not unitary.  

The usage of these irreducible holomorphic representations which are so closely related to the irreps of $\SUtwo$ fits perfectly with our guiding principle of staying close to the description of real LQG and the methods developed there. We will come back to this in later sections. 

Before we show how the holomorphic irreps sit in the general representation theory of $\SLtwoC$, we want to recall their actual form. Again we understand $g\in\SLtwoC$ as a $2\times 2$-matrix as in \eref{eq:g_abcd}. Hence the irreps are
\begin{equation}
    \Pi_{j}(g)_{mn}= \sum_k \frac{\sqrt{(j+m)!(j-m)!(j+n)!(j-n)!}}{(j+n-k)!(m-n+k)!(j-m-k)!k!} a^{j+n-k} b^{m-n+k} c^k d^{j-m-k}, 
\end{equation}
with $j\in\frac{\mathbb{N}}{2}$, $m,n\in\{-j,\dots,j\}$ and $k$ runs over all integers admissible by the factorials. It is a polynomial in the matrix elements of $g$ and therefore obviously holomorphic in accordance to definition \ref{def:analytic_functions}. We further immediately see that 
\begin{equation}
    \overline{\Pi_{j}(g)_{mn}}=\Pi_{j}(\bar{g})_{mn},
\end{equation}
which gives us the anti-holomorphic irreps. 

In the general representation theory of $\SLtwoC$, they are in fact part of the finite dimensional representations. The holomorphic representations are of the type $(j,0)$, while the anti-holomorphic representations are of the type $(0,j)$. This fits perfectly with the property 
\begin{equation}
    \overline{(j_1,j_2)}=(j_2,j_1)
\end{equation}
of the finite dimensional representations. 
%This can be seen as well from \eref{eq:SLtwoC_realisations} in the case of the finite dimensional representations for either $p=0$ or $q=0$ and then only action on polynomials of either $z$ or $\bar{z}$, i.\,e. holomorphic functions. 

Finally we want to remark that an important aspect of LQG in real variables is the decomposition of cylindrical functions in terms of $\SUtwo$ irreps. This is possible because of the Peter-Weyl theorem. Establishing a similar statement for non-compact groups, i.\,e. in our case $\SLtwoC$ is more complicated and manly tied to using the unitary finite dimensional representations. There is however also work on a holomorphic Peter-Weyl theorem for $\SLtwoC$ \cite{huebschmann_kirillovs_2008}. At this point it is  not obvious whether the holomorphic and anti-holomorphic irreps described above can serve a similar purpose here and can be used to fully describe holomorphic and anti-holomorphic functions of $\SLtwoC$. We come back to this in section \ref{sec:ONB}.

\section{Holonomies, fluxes and the quantum algebra}
\label{sec:hfalgebra}

The purpose of this section is to define the algebra of $\SLtwoC$ holonomies and fluxes. We follow the exact steps of real LQG, i.\,e., smearing the Ashtekar connection along paths and the electric field over surfaces, in order to regularise them. 

\subsection{$\SLtwoC$ holonomies and fluxes}
We start with holonomies. For any Lie group $G$, a $G$-connection gives rise to the holonomy along a path $e$ as the path ordered exponential
\begin{equation}
    h_e=\curlyP \exp\curlybrackets{-\int_e\! A},
\end{equation}
with all the usual properties under concatenation and inversion of paths. So compared to $\SUtwo$ holonomies there is no structural difference when we consider the holonomies of an $\SLtwoC$ connection. As $h_e$ now takes values in $\SLtwoC$, we have to consider that here holonomies are not unitary matrices. 

Similarly we construct the flux of the $\SLtwoC$ electric field via integrating it, together with an $\sltwoC$-valued function $f$, over a surface $S$. This yields the usual expression
\begin{equation}\label{eq:flux_def}
    E_f(S)=\int_S (\ast\tE)(f)=\int_S\frac{1}{2}\lc{_{abc}}\tE^a_i f^i \grad{x^b}\wedge\grad{x^c}.
\end{equation}
Compared to the real fluxes of the $\SUtwo$ electric field, we are now dealing with complex fluxes, as both $\tE^a_i$ and $f^i$ take complex values. 

From the Poisson relation \eref{eq:Poisson_Relation_AE} we can obtain the corresponding Poisson relation for $\SLtwoC$ holonomies and fluxes, by following exactly the same procedure for the relation of $\SUtwo$ holonomies and fluxes, as e.\,g. presented in \cite{thiemann_modern_2007}.\footnote{The key point is that the Poisson bracket of holonomies and fluxes in both cases has to be reduced to the ones for connection and electric field. This is done by isolation the latter via a limiting procedure at an intersection point of path and surface. This is a shared feature of both formulations, hence using complex variables does not change anything here.} We do this for a single edge and a single surface and directly consider the commutator via multiplication by $i$. The result is
\begin{equation}
    \commutator{h_e}{E_f(S)}=\frac{\kappa}{4\lambda} \kappa(e,S) \begin{cases}
    h_e f(p) & \text{$p=e\cap S$ source of $e$}\\
     -f(p)  h_e & \text{$p$ target of $e$}
    \end{cases},
\end{equation}
where $\kappa(e,S)$ as usual describes the relative orientation of edge and surface. It is $+1$ or $-1$ if $e$ lies above or below $S$, respectively, and vanishes if there is no intersection or the edge lies in the surface. 
As $f(p)=f^i(p) \tau_i$, we can identify the commutator with the action of the holomorphic left- and right-invariant vector fields \eref{eq:hol_left_vf} and \eref{eq:hol_right_vf} of definition \ref{def:hol_vf}. This turns the commutator into 
\begin{equation}
    \commutator{h_e}{E_f(S)}=\frac{\kappa}{4\lambda} \kappa(e,S) f^i(p) \begin{cases}
    L_i h_e  & \text{$p$ source of $e$}\\
     R_i h_e & \text{$p$ target of $e$}
    \end{cases}.
\end{equation}

With this we obtained two different things. At first, we managed to go over from the holomorphic Poisson bracket, in terms of holomorphic derivatives, to a holomorphic canonical commutation relation, in terms of holomorphic invariant vector fields. 
Secondly, this allows to quantise the complex fluxes exactly like in real LQG. Flux operators act as the commutator with the flux and therefore it is possible to represent them by the action of holomorphic invariant vector fields.

These basic operators and relations are the starting point of a holomorphic holonomy-flux algebra. General cylindrical functions and their relations and commutators with fluxes can be introduced, but since everything works in close analogy to the $\SUtwo$ case, we refrain from spelling out the details here.

\subsection{Comparison to real holonomy-flux-algebra}
Real LQG is the representation theory of the holonomy-flux-algebra. This quantum $\ast$-algebra consists of $\SUtwo$ holonomies of the real Ashtekar-Barbero connection and the flux-operators of the corresponding real electric field. See e.\,g. \cite{lewandowski_uniqueness_2006} for a detailed description.  

Because of the uniqueness result \cite{lewandowski_uniqueness_2006,fleischhack_representations_2009}, the Ashtekar-Lewandowski state \cite{ashtekar_projective_1995} is the only diffeomorphism invariant state on the holonomy-flux-algebra. The resulting Ashtekar-Lewandowski representation is the unique diffeomorphism invariant representation of real LQG. 

As this result has such restricting consequences for LQG, it is a fair question whether this uniqueness theorem possibly  applies for the $\SLtwoC$ quantum algebra described in this section, or not. 

We will answer this on a technical level. An important aspect of the definition of the holonomy-flux-algebra in e.\,g. \cite{lewandowski_uniqueness_2006} is that it is a $\ast$-algebra. It has the adjoint as an involution. This is essential for the uniqueness result.  In what we are doing, it is crucial that we do not have such an involution, as we are trying to construct a quantum theory based on holomorphic functions. Therefore we are not working with a quantum $\ast$-algebra and the uniqueness theorem does not apply for this quantum algebra. In fact, we will see that it is relatively easy to define a whole class of mutually inequivalent, diffeomorphism covariant representations.

\section{Quantum theory and first reality condition}
\label{sec:main}
In this section we present the construction of the quantum theory based on $\SLtwoC$ fluxes and holonomies, while at the same implementing the first reality condition. We start with an overview of our approach.

\subsection{General strategy} 
Using the holomorphic toolbox introduced in section \ref{sec:analytic_functions} our objective is to formulate a holomorphic quantum theory, structurally close to real LQG. 
Expressed already in terms of $\SLtwoC$ elements and holomorphic invariant vector fields and regularised with paths and surfaces, holonomies and fluxes are already of the desired form. 

The general understanding in the literature, cf. \cite{ashtekar_self_1992,tate_algebraic_1993,ashtekar_lectures_1998,wilson-ewing_loop_2015}, is that the reality conditions are supposed to be implemented as adjointness relations, which are additional conditions in the construction of the quantum theory. 
The classical reality conditions from section \ref{sec:classicaltheory}, therefore, have to be carefully turned into suitable adjointness relations for holonomies and fluxes. 

An implementation as adjointness relations means adjointness with respect to some inner product. In this sense the reality conditions are supposed to be implemented via the choice of an appropriate inner product for the quantum theory.

Further, we want to use the choice of an inner product to tame the non-compactness of $\SLtwoC$. As this is closely related to the matter of cylindrical consistency of a measure, we will come back to this only later on. 

As it turns out, it is beneficial to simultaneously develop the quantum theory and implement the first reality condition. 
For this reason, we only consider RCI here and refer to \cite{sahlmann_revisiting_RCII} for a treatment of the second reality condition.

First we need to formulate an adjointness relation for flux operators, which corresponds to the reality of the electric field.
The strategy then is to start with the smallest structures used in LQG quantisation, namely the Hilbert space corresponding to a single edge. On this we want to define an inner product which leaves enough freedom in the choice of a measure, in order to implement the first reality condition. 

From this smallest building block we then recover well known structures  from real LQG, but adapted to our situation. In particular, an $\SLtwoC$ spin network basis and a class of consistent measures on the holomorphic Hilbert space of generalised connections.

\subsection{Adjointness relations for fluxes and invariant vector fields}
\label{sec:adjointness}
Before we start the construction of the actual quantum theory, we need to turn the reality condition for the electric field into an adjointness relation for fluxes. 

We consider the flux \eref{eq:flux_def} without smearing function, i.\,e. 
\begin{equation}
    E_j(S)=\int_S\frac{1}{2}\lc{_{abc}}\tE^a_j  \grad{x^b}\wedge\grad{x^c}.
\end{equation}
Therefore reality of the electric field implies reality of the flux $E_j(S)=\overline{E_j(S)}$. 
We understand the electric field as an object living in the dual space of $\sltwoC$. Hence  
\begin{equation}
    E_j(S)\equiv E_{\tau_j}(S)= \int_S\frac{1}{2}\lc{_{abc}}\tE^a_i \tau^{\ast i}(\tau_j)  \grad{x^b}\wedge\grad{x^c}.
\end{equation}
This allows to deduce the reality of fluxes including smearing with respect to an $\sltwoC$-valued function $f(x)=f^j(x)\tau_j$. It yields the condition
\begin{equation}
    E_f(S)=-\overline{E_{f^\dagger}(S)}.
\end{equation}
We use the same notation $E_j(S)$ and $E_f(S)$ for the actual flux operators in the following. With respect to the Hilbert space adjoint, the reality condition for fluxes takes the equivalent forms
\begin{equation}\label{eq:RCI_flux}
    {E_j(S)}^\dagger=E_j(S)
\end{equation}
and
\begin{equation}
    {E_f(S)}^\dagger=-E_{f^\dagger}(S).
\end{equation}

We already know how to represent the flux operators in terms of holomorphic invariant vector fields. Therefore we can reformulate the reality condition as a condition on them. Staying in the picture of a single edge, the flux operator is given by 
\begin{equation}
    E_j(S)\Psi(h_e)=-\frac{\kappa}{4\lambda} \kappa(e,S) J_j \Psi(h_e),
\end{equation}
where $J_i$ is the left- or right-invariant holomorphic vector field, depending on whether the intersection point is the source or target of the edge. Here $\Psi(h_e)$ is a holomorphic function of the holonomy in the sense of definition \ref{def:analytic_functions}.

Equation \eref{eq:RCI_flux} now tells us that the reality condition is satisfied if
\begin{equation}
    E_j(S)\Psi(h_e)=E_j(S)^\dagger\Psi(h_e)=-\frac{\kappa}{4} \kappa(e,S) \rb{\frac{1}{\lambda}J_j}^\dagger \Psi(h_e),
\end{equation}
which tells us that the combination of $\lambda$ and the invariant holomorphic vector fields has to be selfadjoint:
\begin{equation}\label{eq:RCI_invariant_vector_fields}
    \rb{\frac{1}{\lambda}J_j}^\dagger=\frac{1}{\lambda}J_j.
\end{equation}
Recall that we are interested in the special cases $\lambda=1,i$. So the invariant vector fields are supposed to be either self- or skew-adjoint. 

\subsection{Single edge Hilbert space and measure ansatz}
Recall that we want to start with the Hilbert for a single edge, represented by the corresponding holonomy, taking values in $\SLtwoC$. It is supposed to consist of holomorphic functions in accordance with definition \ref{def:analytic_functions}. Therefore we look at 
\begin{equation}
    \curlyH:=\curlyL^2\rb{\SLtwoC,\grad{g_\mathrm{H}}\,\mu(g,\bar{g})}\cap \mathfrak{H}(\SLtwoC).
\end{equation}
Functions in $\curlyH$ are therefore both square integrable and  holomorphic. Further $\grad{g_\mathrm{H}}$ is the Haar-measure on $\SLtwoC$. In order that $\grad{g_\mathrm{H}}\,\mu(g,\bar{g})$ is a measure, the function $\mu(g,\bar{g})$ has to be continuous and non-negative. It is added to tame the non-compactness of $\SLtwoC$ and is supposed to be determined by this requirement and the reality conditions. Restrictions on its form are therefore specified later on. 
For $\Phi,\Psi\in\curlyH$, the inner product takes the form 
\begin{equation}\label{eq:inner_product_one_edge_HS}
    \innerproduct{\Phi}{\Psi}=\int_{\SLtwoC}\!\grad{g_\mathrm{H}}\,\mu(g,\bar{g})\, \overline{\Phi(g)}\Psi(g).
\end{equation}
There are several things to remark here. While $\Phi(g)$ is a holomorphic function, its complex conjugate $\overline{\Phi(g)}$ is necessarily anti-holomorphic. So even though we want to construct a holomorphic quantum theory, there are anti-holomorphic objects present, although only in the inner product, which is not a function of $\SLtwoC$ and therefore does not even have to be holomorphic. While $\Psi(g)$ and $\Phi(g)$ are either holomorphic or anti-holomorphic, the additional function $\mu(g,\bar{g})$ is not of either form. This is relevant for obtaining a finite inner product and therefore taming the non-compactness of $\SLtwoC$. Because it allows for combinations of $g$ and its complex conjugate, it gives individual access to the rotation and boost parts of the group element. This should in principle allow to tame the non-compactness,  caused by the boost part, with an appropriate fall off behaviour of the measure in these directions.

\subsection{Reality condition for fluxes}
\label{sec:realitycondition}
The next step is to turn the first reality condition, in the form of \eref{eq:RCI_invariant_vector_fields}, into a condition on the measure function $\mu(g,\bar{g})$. We have to do this for left- and right-invariant vector fields individually. Starting with the left-invariant case, we find  
\begin{equation}
    \begin{aligned}
        \innerproduct{\Phi}{L_i\Psi}&=\int_{\SLtwoC}\!\grad{g_\mathrm{H}}\mu(g,\bar{g}) \overline{\Phi(g)} L_i\Psi(g)\\
        &=\int_{\SLtwoC}\!\grad{g_\mathrm{H}}\mu(g,\bar{g}) \overline{\Phi(g)} \frac{\grad{}}{\grad{}\varepsilon}
        \Big\vert_{\varepsilon=0}\Psi(g e^{\varepsilon\tau_i}).
    \end{aligned}
\end{equation}
Using the Leibniz rule, we move the $\varepsilon$-derivative to the left, as everything besides $\Psi$ is independent of $\varepsilon$. Further we use the invariance of the Haar-measure to shift the right-translation of $\Psi$ to every other term. Hence
\begin{equation}
    \begin{aligned}
       \innerproduct{\Phi}{L_i\Psi}&=\int_{\SLtwoC}\!\grad{g_\mathrm{H}}\frac{\grad{}}{\grad{}\varepsilon}
        \Big\vert_{\varepsilon=0} \mu(g,\bar{g}) \overline{\Phi(g)} \Psi(g e^{\varepsilon\tau_i}) \\
        &=\int_{\SLtwoC}\!\grad{g_\mathrm{H}}\frac{\grad{}}{\grad{}\varepsilon}
        \Big\vert_{\varepsilon=0} \mu\rb{ge^{-\varepsilon\tau_i},\overline{ge^{-\varepsilon\tau_i}}} \overline{\Phi(ge^{-\varepsilon\tau_i})} \Psi(g).
    \end{aligned}
\end{equation}
As the $\varepsilon$-derivative is holomorphic, the derivative of $\overline{\Phi(ge^{-\varepsilon\tau_i})}$ as a function of $\bar{\varepsilon}$ vanishes. We set $\varepsilon$ to zero in this contribution, but keep it in the second argument of the measure function:
\begin{equation}\label{eq:RCI_aux_1}
    \innerproduct{\Phi}{L_i\Psi}=\int_{\SLtwoC}\!\grad{g_\mathrm{H}}\frac{\grad{}}{\grad{}\varepsilon}
        \Big\vert_{\varepsilon=0} \mu\rb{ge^{-\varepsilon\tau_i},\overline{ge^{-\varepsilon\tau_i}}} \overline{\Phi(g)} \Psi(g).
\end{equation}
Further we need the adjoint of the left-invariant vector fields. For this we use the very definition of the adjoint:
\begin{equation}
\begin{aligned}
    \innerproduct{\Phi}{(L_i)^\dagger\Psi}&=\innerproduct{L_i\Phi}{\Psi}=\int_{\SLtwoC}\!\grad{g_\mathrm{H}}\mu(g,\bar{g}) \overline{L_i\Phi(g)} \Psi(g)\\
    &=\int_{\SLtwoC}\!\grad{g_\mathrm{H}}\mu(g,\bar{g}) \overline{\frac{\grad{}}{\grad{}\varepsilon}
        \Big\vert_{\varepsilon=0}\Phi(ge^{\varepsilon\tau_i})} \Psi(g).
\end{aligned}    
\end{equation}
The complex conjugate of the holomorphic derivative is the anti-holomorphic derivative. We use again the Leibniz rule to move the derivative and the invariance of the Haar-measure to shift the right-translation from $\Phi$ to the measure, as $\Psi$ becomes a function of $\varepsilon$, annihilated by the $\bar{\varepsilon}$-derivative. This results in 
\begin{equation}\label{eq:RCI_aux_2}
    \innerproduct{\Phi}{(L_i)^\dagger\Psi}=\int_{\SLtwoC}\!\grad{g_\mathrm{H}}\frac{\grad{}}{\grad{}\bar{\varepsilon}}
        \Big\vert_{\bar{\varepsilon}=0} \mu\rb{ge^{-\varepsilon\tau_i},\overline{ge^{-\varepsilon\tau_i}}} \overline{\Phi(g)} \Psi(g).
\end{equation}

The actual adjointness relation \eref{eq:RCI_invariant_vector_fields} is $\lambda$-dependent. Taking this into account, when combining \eref{eq:RCI_aux_1} and \eref{eq:RCI_aux_2}  into the reality condition, we find 
\begin{equation}
\label{eq:rci_aux}
    \innerproduct{\Phi}{L_i\Psi}=\frac{\lambda}{\bar{\lambda}} \innerproduct{\Phi}{(L_i)^\dagger\Psi}.
\end{equation}
For the cases $\lambda=1,i$, the left-invariant vector field has to be either selfadjoint ($\lambda=1$) or skew-adjoint ($\lambda=i$):\footnote{We could in principle also consider $\lambda$ to be a general phase, however, the conditions \eref{eq:rci_aux} imposed by RCI look much less natural and would be hard to satisfy.}
\begin{equation}
    \innerproduct{\Phi}{L_i\Psi}=\pm \innerproduct{\Phi}{(L_i)^\dagger\Psi}.
\end{equation}
As we transferred the action of the invariant vector field to the measure function only, this is equivalent to 
\begin{equation}
    \frac{\grad{}}{\grad{}\varepsilon}
        \Big\vert_{\varepsilon=0} \mu\rb{ge^{\varepsilon\tau_i},\overline{ge^{\varepsilon\tau_i}}}=\pm \frac{\grad{}}{\grad{}\bar{\varepsilon}}
        \Big\vert_{\bar{\varepsilon}=0} \mu\rb{ge^{\varepsilon\tau_i},\overline{ge^{\varepsilon\tau_i}}},
\end{equation}
where we changed to sign of $\varepsilon$ in the exponents at no cost. The occurring sign when changing the $\varepsilon$- and $\bar{\varepsilon}$-derivatives cancel.
Rewriting this as a derivative of one measure function only, yields
\begin{equation}\label{eq:RCI_aux_3}
    \rb{\frac{\grad{}}{\grad{}\varepsilon}
         \mp \frac{\grad{}}{\grad{}\bar{\varepsilon}}}
        \Big\vert_{\varepsilon=0} \mu\rb{ge^{\varepsilon\tau_i},\overline{ge^{\varepsilon\tau_i}}} =0.
\end{equation}
Depending on the sign, i.\,e., the value of $\lambda$, we can identify the combination of holomorphic and anti-holomorphic derivatives as the boost and rotational derivatives of equations \eref{eq:rotational_and_boost_derivatives} and \eref{eq:rotational_and_boost_invariant_vector_fields}.

Before we further analyse this, we want to derive the corresponding equation for right-invariant vector fields. Starting from 
\begin{equation}
    \innerproduct{\Phi}{R_i\Psi}=\pm \innerproduct{\Phi}{(R_i)^\dagger\Psi},
\end{equation}
virtually identical steps have to be performed, in order to transfer the holomorphic and anti-holomorphic derivatives to the measure function, with the main difference that we are dealing with the left-action instead of the right-action. As a condition on the measure function we therefore find
\begin{equation}\label{eq:RCI_aux_4}
    \rb{\frac{\grad{}}{\grad{}\varepsilon}
         \mp \frac{\grad{}}{\grad{}\bar{\varepsilon}}}
        \Big\vert_{\varepsilon=0} \mu\rb{e^{\varepsilon\tau_i}g,\overline{e^{\varepsilon\tau_i}g}} =0.
\end{equation}
Again, the $\varepsilon$- and $\bar{\varepsilon}$-derivatives form the rotational and boost derivatives, but now with respect to the right-action. Irrespective of the type of invariant vector field, \eref{eq:RCI_aux_3} and \eref{eq:RCI_aux_4} yield the same condition on the measure function in order to implement reality of flux operators. For ``$-$'' in \eref{eq:RCI_aux_3} and \eref{eq:RCI_aux_4}, i.\,e. $\lambda=1$, the boost derivative must vanish. Therefore $\mu(g,\bar{g})$ has to only be a function of the rotational part of $g$. Contrary to this, ``$+$'' in \eref{eq:RCI_aux_3} and \eref{eq:RCI_aux_4}, i.\,e. $\lambda=i$, requires the rotational derivative to vanish. Hence $\mu(g,\bar{g})$ is only depending on the boost part of $g$. 

Therefore, we obtain two different classes of measures, that are both able to implement the reality of fluxes. The ones that are rotationally invariant and the ones that are invariant under boosts. As described earlier, by the choice of measure, we want to solve both, the reality condition and the problem of non-compactness of $\SLtwoC$ at the same time. In terms of the measure, this requires the measure function $\mu(g,\bar{g})$ to tame the integration in the boost directions. This requires $\mu(g,\bar{g})$ to be a function of the boost part of $g$ and tells us that the boost invariant measure function, associated with the case $\lambda=1$ is not suitable for this approach. Here we do not have any control over the non-compact boost directions in the measure. 
This means our approach favours $\lambda=i$. 

Before we go on with formulating a measure solving RCI in the next section, we want to comment on the $\lambda=1$ case. In \cite{wilson-ewing_loop_2015} loop quantum cosmology is considered in a selfdual formulation. The reality conditions (both at the same time) are solved by the choice of a distributional measure, which is only a function of the rotational parameter. It solves the first reality condition but fails to tame the non-compactness of, in this case, the complexification of $\Uone$. This is however overcome because LQC and its selfdual counterpart use the Bohr-compactification of the real line. This results in a finite inner product. 
As there is no Bohr-compactification of $\SLtwoC$, using similar methods as in \cite{wilson-ewing_loop_2015} is not possible here.\footnote{The construction of the Bohr compactification uses a mean on the group. It can be shown that a mean does not exist for $\SLtwoC$.}

\subsection{Solving the first reality condition}
\label{sec:solution}
We consider the case $\lambda=i$.
Although \eref{eq:RCI_aux_3} and \eref{eq:RCI_aux_4}, for ``$+$'', both demand rotational invariance, they are still two separate conditions. One with respect to the right-action and one with respect to the left-action. Therefore, a solution for the reality condition has to account for both of these situations. 

Rotational invariance specifically refers to invariance with respect to the rotational subgroup of $\SLtwoC$ and therefore $\SUtwo$. 
This suggests to look at the polar decomposition of $g\in\SLtwoC$ into a positive semi-definite Hermitian matrix $|g|_\mathrm{L}$ and a unitary matrix $U_\mathrm{L}$:
\begin{equation}\label{eq:Polar_Dec_L}
    g=|g|_\mathrm{L}U_\mathrm{L} \quad \text{with}\quad |g|_\mathrm{L}=\sqrt{gg^\dagger} \quad \text{and}\quad U_\mathrm{L}=(|g|_\mathrm{L})^{-1} g.
\end{equation}
This decomposition however is not unique, as one can also decompose 
\begin{equation}\label{eq:Polar_Dec_R}
    g=U_\mathrm{R}|g|_\mathrm{R} \quad \text{with}\quad |g|_\mathrm{R}=\sqrt{g^\dagger g} \quad \text{and}\quad U_\mathrm{R}=g(|g|_\mathrm{R})^{-1}.
\end{equation}
Consequently $|g|_{\mathrm{L,R}}$ represents the boost part of $g$, while $U_{\mathrm{L,R}}\in\SUtwo$ are rotations. 

Invariance under $\SUtwo$ implies that the measure function is, in fact, only a function of $|g|$. At this point it becomes relevant that we have indeed two equations, \eref{eq:RCI_aux_3} and \eref{eq:RCI_aux_4}, that differentiate between the right- and left-action of rotations. For $\alpha\in\RR$, $|ge^{\alpha\tau_i}|_\mathrm{L}$ is obviously invariant under the rotation acting from the right, while $|e^{-\alpha\tau_i}g|_\mathrm{R}$ is invariant under the rotation acting from the left. However, since
\begin{equation}
    |e^{-\alpha\tau_i}g|_\mathrm{L}=\sqrt{e^{-\alpha\tau_i}gg^\dagger e^{\alpha\tau_i}}\quad\text{and}\quad |ge^{\alpha\tau_i}|_\mathrm{R}=\sqrt{e^{-\alpha\tau_i}g^\dagger g e^{\alpha\tau_i}}
\end{equation}
are not invariant under left- and right-action, respectively, it is not possible to choose just one of them as the argument of the measure function. The solution to this is the incorporation of group averaging. We look at 
\begin{equation}\label{eq:group_av_aux_1}
    \int_\SUtwo\!\grad{h_H} f(|hg|_\mathrm{L})= \int_\SUtwo\!\grad{h_\mathrm{H}} f\rb{\sqrt{hgg^\dagger h^{-1}}}.
\end{equation}
While the right-action of $\exp(\alpha\tau_i)$ cancels as before, the left-action of $\exp(-\alpha\tau_i)$, an $\SUtwo$-element, can be absorbed into the invariant Haar-measure of $\SUtwo$. The procedure for group averaging $|g|_\mathrm{R}$ works analogously, but the averaging has to be performed with respect to the right-action of $h$:
\begin{equation}\label{eq:group_av_aux_2}
    \int_\SUtwo\!\grad{h_H} f(|gh|_\mathrm{R})= \int_\SUtwo\!\grad{h_\mathrm{H}} f\rb{\sqrt{hg^\dagger gh^{-1}}}.
\end{equation}

Note that $g$ and $g^\dagger$ do not commute in general. In the decomposed forms \eref{eq:Polar_Dec_L} and \eref{eq:Polar_Dec_R}, however, we see 
\begin{equation}
    \begin{aligned}
        gg^\dagger&= U_\mathrm{R}|g|_\mathrm{R}|g|_\mathrm{R} {U_\mathrm{R}}^{-1}=U_\mathrm{R}g^\dagger g {U_\mathrm{R}}^{-1},\\
        g^\dagger g&= {U_\mathrm{L}}^{-1}|g|_\mathrm{L}|g|_\mathrm{L} {U_\mathrm{L}}={U_\mathrm{L}}^{-1}gg^\dagger  {U_\mathrm{L}}.
    \end{aligned}
\end{equation}
Hence, they commute up to unitary transformations, which can be absorbed by group averaging. As a consequence, we find
\begin{equation}\label{eq:group_av_aux_3}
    \int_\SUtwo\!\grad{h_\mathrm{H}} f\rb{\sqrt{hgg^\dagger h^{-1}}} = \int_\SUtwo\!\grad{h_\mathrm{H}} f\rb{\sqrt{hg^\dagger gh^{-1}}}.
\end{equation}

Having established this group averaging of $|g|$, we can formulate a
 general solution of the first reality condition in the case $\lambda=i$. It is given by measure functions of the form 
\begin{equation}\label{eq:RCI_measure_function}
    \mu(g,\bar{g})=\int_\SUtwo\!\grad{h_H} \mu(hgg^\dagger h^{-1}),
\end{equation}
such that $\mu(g,\bar{g})$ is continuous and non negative. If we chose $\mu$ continuous and non-negative, then so will be $\mu(g,\bar{g})$: since the group averaging is with respect to a positive measure, the result is again a positive function. Since the group averaging is given by an integral over a compact set, it can not produce a non-continuous result. 

From what we have established, it is clear that \eref{eq:RCI_measure_function} satisfies \eref{eq:RCI_aux_3} and \eref{eq:RCI_aux_4} in the case $\lambda=1$. 

Several remarks are in order. Without the symmetry in $g$ and $g^\dagger$, enabled by \eref{eq:group_av_aux_3}, we would not have the single solution \eref{eq:RCI_measure_function} for the first reality condition. In contrast, this would result in a variety of non-convertible measure functions with respect to the ordering of $g$ and $g^\dagger$, like $gg^\dagger$, $g^\dagger g$, $gg^\dagger+g^\dagger g$ or combinations thereof. All of them would satisfy the reality condition independently, as the $gg^\dagger$ to $g^\dagger g$ symmetry is nothing we need for the reality condition. Nevertheless it simplifies the solution considerably. 

With a solution of the type \eref{eq:RCI_measure_function} in mind, we recall the inner product \eref{eq:inner_product_one_edge_HS} on the holomorphic Hilbert space $\curlyH$. It is characterised by the measure
\begin{equation}\label{eq:RCI_measure}
    \grad{g_\mathrm{H}}\mu(g,\bar{g})=\grad{g_\mathrm{H}}\int_\SUtwo\!\grad{h_\mathrm{H}}\mu(hgg^\dagger{h}^{-1}).
\end{equation}
For $g\in\SLtwoC$ and $h\in\SUtwo$, the Haar-measures $\grad{g_\mathrm{H}}$ and $\grad{h_\mathrm{H}}$ are the invariant measures for the respective groups. With the inclusion of $\mu(g,\bar{g})$ as in \eref{eq:RCI_measure_function}, by construction, $\grad{g_\mathrm{H}}\mu(g,\bar{g})$ is only invariant under $\SUtwo$. The measure on $\curlyH$ and therefore the inner product is not invariant under $\SLtwoC$. We will come back to this once we consider gauge invariance and the implementation of the Gauß constraint in section \ref{sec:spin_networks_gauge_invariance}.

\subsection{Holomorphic irreps in the context of RCI}\label{sec:ONB}
The purpose of this section is to analyse the inner product of holomorphic irreducible representations of $\SLtwoC$, as described in section \ref{sec:holomorphic_irreps}, with respect to the measure \eref{eq:RCI_measure_function}, which implements the reality condition. 

Under polar decomposition the Haar-measure of $\SLtwoC$ takes the form 
\begin{equation}\label{eq:Haar_measure_polar_decomp}
    \grad{g_\mathrm{H}}=\grad{|g|}\grad{U_\mathrm{H}},
\end{equation}
as it is shown in \cite{hall_phase_1997}.
Here $\grad{U_\mathrm{H}}$ is the Haar-measure on the compact rotation subgroup $\SUtwo$. This is valid for both types of polar decomposition \eref{eq:Polar_Dec_L} and \eref{eq:Polar_Dec_R}, which is why we do not differentiate these cases in \eref{eq:Haar_measure_polar_decomp}. Because we do not use the actual parametrisation for explicit integrations, we refer to \cite{hall_phase_1997} and refrain from a detailed description\footnote{In \cite{hall_phase_1997} $\SLtwoC$ is parameterised by an $\SUtwo$ element, call it $U$, and $\exp(ia)$, where $a$ is an $\sutwo$ element. The latter takes on the role of $|g|$. The integration is over the $\SUtwo$ Haar measure and a measure $\sigma(a)\grad{a}$, where $\sigma(a)$ is a function of $\mathrm{ad}_a$. The measure decomposition is shown for one of the orders in the polar decomposition, but can be repeated for the other order as well.}.
We can decompose $g\in\SLtwoC$ in terms of \eref{eq:Polar_Dec_L} or \eref{eq:Polar_Dec_R}. Therefore the irreps split into 
\begin{equation}
\begin{aligned}
    \IR{j}{g}{^m_n}&=\IR{j}{|g|_\mathrm{L}U_\mathrm{L}}{^m_n}=\IR{j}{|g|_\mathrm{L}}{^m_l}\IR{j}{U_\mathrm{L}}{^l_n}\\
    &=\IR{j}{U_\mathrm{R}|g|_\mathrm{R}}{^m_n}=\IR{j}{U_\mathrm{R}}{^m_l}\IR{j}{|g|_\mathrm{R}}{^l_n},
    \end{aligned}
\end{equation}
as $|g|_\mathrm{L/R}\in\SLtwoC$ and $U_\mathrm{L/R}\in\SLtwoC$.
We choose to work in the ``L''-decomposition for now. Already here we want to state that holomorphic irreps of $\SLtwoC$ reduce to the standard irreps of $\SUtwo$ when we only consider $U\in\SUtwo$. Therefore the orthogonality relation 
\begin{equation}
    \int_\SUtwo\!\grad{U_\mathrm{H}}\overline{\IR{j_1}{U}{^{m_1}_{n_1}}}\IR{j_2}{U}{^{m_2}_{n_2}}=\frac{1}{2j_1+1}\delta_{j_1j_2}\delta^{m_1m_2}\delta_{n_1n_2}
\end{equation}
holds. Further, as $\mu(g,\bar{g})$ is effectively a function of $gg^\dagger$ and invariant under the order of $g$ and $g^\dagger$, we denote it as $\mu(|g|^2)$.  The inner product of irreps is given by
\begin{equation}
    \begin{aligned}
        \innerproduct{\tensor{{\Pi_{j_1}}}{^{m_1}_{n_1}}}{\tensor{{\Pi_{j_2}}}{^{m_2}_{n_2}}}&=\int\grad{|g|}\grad{U_\mathrm{H}}\mu(|g|^2) \overline{\IR{j_1}{|g|_\mathrm{L}U_\mathrm{L}}{^{m_1}_{n_1}}}\IR{j_2}{|g|_\mathrm{L}U_\mathrm{L}}{^{m_2}_{n_2}}\\
        &=\int\grad{|g|}\grad{U_\mathrm{H}}\mu(|g|^2) \overline{\IR{j_1}{|g|_\mathrm{L}}{^{m_1}_{l_1}} \IR{j_1}{U_\mathrm{L}}{^{l_1}_{n_1}}} \IR{j_2}{|g|_\mathrm{L}}{^{m_2}_{l_2}} \IR{j_2}{U_\mathrm{L}}{^{l_2}_{n_2}}\\
        &=\int\grad{|g|}\mu(|g|^2)\overline{\IR{j_1}{|g|_\mathrm{L}}{^{m_1}_{l_1}}} \IR{j_2}{|g|_\mathrm{L}}{^{m_2}_{l_2}} \frac{1}{2j_1+1}\delta_{j_1j_2}\delta^{l_1l_2}\delta_{n_1n_2}\\
        &=\frac{1}{2j_1+1}\delta_{j_1j_2}\delta_{n_1n_2}\int\grad{|g|}\mu(|g|^2)\overline{\IR{j_1}{|g|_\mathrm{L}}{^{m_1l_2}}} \IR{j_1}{|g|_\mathrm{L}}{^{m_2}_{l_2}},
    \end{aligned}
\end{equation}
where we used the orthogonality of $\SUtwo$ irreps from the second to the third line. With respect to the matrix adjoint and using that $|g|_\mathrm{L}$ is selfadjoint, we see 
\begin{equation}
    \overline{\IR{j_1}{|g|_\mathrm{L}}{^{m_1l_2}}}=\IR{j_1}{(|g|_\mathrm{L})^\dagger}{^{l_2m_1}}=\IR{j_1}{|g|_\mathrm{L}}{^{l_2m_1}},
\end{equation}
which allows to combine the irreps left in the inner product.
Therefore we are left with 
\begin{equation}\label{eq:ONB_aux_1}
    \innerproduct{\tensor{{\Pi_{j_1}}}{^{m_1}_{n_1}}}{\tensor{{\Pi_{j_2}}}{^{m_2}_{n_2}}}=\frac{1}{2j_1+1}\delta_{j_1j_2}\delta_{n_1n_2}\int\grad{|g|}\mu(|g|^2) \IR{j_1}{{|g|_\mathrm{L}}^2}{^{m_2m_1}}. 
\end{equation} 

Performing a virtually identical calculation, we can look at the ``R''-decomposition of the inner product. Namely
\begin{equation}\label{eq:ONB_aux_2}
    \begin{aligned}
        \innerproduct{\tensor{{\Pi_{j_1}}}{^{m_1}_{n_1}}}{\tensor{{\Pi_{j_2}}}{^{m_2}_{n_2}}}&=\int\grad{|g|}\grad{U_\mathrm{H}}\mu(|g|^2) \overline{\IR{j_1}{U_\mathrm{R}|g|_\mathrm{R}}{^{m_1}_{n_1}}}\IR{j_2}{U_\mathrm{R}|g|_\mathrm{R}}{^{m_2}_{n_2}}\\
        &=\int\grad{|g|}\grad{U_\mathrm{H}}\mu(|g|^2) \overline{\IR{j_1}{U_\mathrm{R}}{^{m_1}_{l_1}} \IR{j_1}{|g|_\mathrm{R}}{^{l_1}_{n_1}}} \IR{j_2}{U_\mathrm{R}}{^{m_2}_{l_2}} \IR{j_2}{|g|_\mathrm{R}}{^{l_2}_{n_2}}\\
        =&\frac{1}{2j_1+1}\delta_{j_1j_2}\delta^{m_1m_2}\int\grad{|g|}\mu(|g|^2) \IR{j_1}{{|g|_\mathrm{L}}^2}{_{n_1n_2}}.
    \end{aligned}
\end{equation}

When we now combine the results \eref{eq:ONB_aux_1} and \eref{eq:ONB_aux_2}, we realise that the inner product is of the form 
\begin{equation}\label{eq:ONB_aux_3}
    \innerproduct{\tensor{{\Pi_{j_1}}}{^{m_1}_{n_1}}}{\tensor{{\Pi_{j_2}}}{^{m_2}_{n_2}}}=c_{j_1}\delta_{j_1j_2}\delta^{m_1m_2}\delta_{n_1n_2},
\end{equation}
with a set of yet to be determined constants $c_{j_1}$. It can be obtained from taking the traces with respect to the magnetic quantum numbers in \eref{eq:ONB_aux_3}, i.\,e. 
\begin{equation}
\begin{aligned}
    c_{j_1}&=\frac{1}{(2j+1)^2}\delta_{m_1m_2}\delta^{n_1n_2} \innerproduct{\tensor{{\Pi_{j_1}}}{^{m_1}_{n_1}}}{\tensor{{\Pi_{j_1}}}{^{m_2}_{n_2}}}\\
    &=\frac{1}{2j_1+1} \int\grad{|g|}\mu(|g|^2) \IR{j_1}{{|g|_\mathrm{L/R}}^2}{^{n_1}_{n_1}}.
\end{aligned}    
\end{equation}
Accordingly, the inner product of holomorphic irreducible representations yields the orthogonality relation 
\begin{equation}\label{eq:Hol_irreps_orthog}
    \innerproduct{\tensor{{\Pi_{j_1}}}{^{m_1}_{n_1}}}{\tensor{{\Pi_{j_2}}}{^{m_2}_{n_2}}}=\frac{\mu_{j_1}}{2j_1+1}\delta_{j_1j_2}\delta^{m_1m_2}\delta_{n_1n_2}.
\end{equation}
Here we introduced the constants $\mu_j$, which, in following, will be referred to as moments of the measure \eref{eq:RCI_measure_function}. They are given by
\begin{equation}
    \mu_j=\int\grad{|g|}\mu(|g|^2) \tr\rb{\IR{j}{{|g|}^2}{}}.
\end{equation}

As we stated already earlier, the Peter-Weyl theorem for $\SUtwo$, which essentially tells us that the normalised $\SUtwo$-irreps serve as an orthonormal basis of $\curlyL^2(\SLtwoC,\grad{h}_\mathrm{H})$, does not apply for the holomorphic irreps of $\SLtwoC$. Nevertheless we can normalise these, i.\,e.
\begin{equation}
    \bIR{j}{g}{^m_n}=\sqrt{\frac{2j+1}{\mu_j}}\IR{j}{g}{^m_n}
\end{equation}
and therefore \eref{eq:Hol_irreps_orthog} turns into 
\begin{equation}
    \innerproduct{\tensor{{b_{j_1}}}{^{m_1}_{n_1}}}{\tensor{{b_{j_2}}}{^{m_2}_{n_2}}}=\delta_{j_1j_2}\delta^{m_1m_2}\delta_{n_1n_2}.
\end{equation}
The set of $\bIR{j}{g}{}$ forms an orthonormal system with respect to the RCI implementing inner product. Recall that we can express every analytic function of $\SLtwoC$  in terms of the holomorphic irreps. With this, the normalised irreps $\bIR{j}{g}{}$ indeed serve as an orthonormal basis of the holomorphic $\curlyL^2$-space $\curlyH$. In particular we found an analogue result to the Peter-Weyl theorem for $\SLtwoC$. 

We want to take a look at what contributed to this. At first, the reality condition for $\lambda=i$ led to a measure function, that is only depending on the boost part of $\SLtwoC$. This allowed to split the $\SUtwo$ integration from the inner product and to use the orthogonality relation. This in turn is only possible because we started out with a holomorphic quantum theory, which led to consider the holomorphic irreps of $\SLtwoC$ and their reduction $\SUtwo$-irreps. In this sense the $\SLtwoC$ orthogonality is inherited from the orthogonality of the subgroup $\SUtwo$. 

Alternatively, if we would have considered the reality condition for $\lambda=1$, the measure function would have to be boost invariant and therefore only a function of $\SUtwo$. The orthogonality of $\SUtwo$ irreps is with respect to the Haar-measure, not with respect to an additional function. Therefore, it would not be possible to employ the orthogonality relation and we would not get an orthogonality relation for $\SLtwoC$-irreps. Because of this it is not obvious if a similar structure would emerge when looking at $\lambda=1$. 

In this sense implementing the first realty condition as we did, enables the existence of the orthonormal basis of holomorphic irreps.  

\subsection{Cylindrical consistency}
\label{sec:consistency}
In the last section we showed that the holomorphic $\SLtwoC$-irreps in our work play a virtually identical role as the $\SUtwo$-irreps in real LQG. Now recall that a characteristic feature of real LQG is the use spin network functions with respect to a given graph $\gamma$, which in span the Hilbert space of generalised connections with respect to a cylindrically consistent measure. 
In this section we want to construct such a measure for selfdual LQG.

We consider a graph $\gamma$, consisting of edges $e_i$ and the denote the set of edges as $E(\gamma)$. A holomorphic function 
\begin{equation}
    F:\curlyA\rightarrow\CC,
\end{equation}
where $\curlyA$ is the space of smooth $\SLtwoC$ connections, is called cylindrical with respect to $\gamma$ in the usual sense, i.\,e. there exists a function
\begin{equation}
    F_\gamma:\SLtwoC^{|E(\gamma)|}\rightarrow\CC,
\end{equation}
such that for $A\in\curlyA$ 
\begin{equation}
    F(A)=F_\gamma\rb{\curlybrackets{h_e(A)}_{e\in E(\gamma)}}.
\end{equation}
We use a product measure to act on such cylindrical functions. This allows us to use the measure \eref{eq:RCI_measure} for all individual edges.  We use the notation 
\begin{equation}
    \grad{\mu(g_e)}=\grad{g_{e\mathrm{H}}}\mu{\rb{g_e{g_e}^\dagger}}
\end{equation}
for the RCI implementing measure on the Hilbert space corresponding to $e\in E(\gamma)$. Note that we suppress the group averaging here, in order to keep the notation compact. Having established this, we define
\begin{equation}\label{eq:cylcon_measure}
    \mu(F)=\int_{\SLtwoC^{|E(\gamma)|}}\prod_{e\in E(\gamma)} \grad{\mu(g_e)} F_\gamma\rb{\curlybrackets{g_e}_{e\in E(\gamma)}}.
\end{equation}
In order for this to be a viable measure for the holomorphic Hilbert space of generalised connections, which we want to construct here, it has to satisfy certain consistency conditions. They ensure that 
\begin{equation}
    \mu(F_\gamma)=\mu(F_{\gamma'})
\end{equation}
whenever $\gamma$ can be embedded in $\gamma'$, or arises from inversion or composition of edges. 
It is then called cylindrically consistent measure.\footnote{For a detailed description in the case of real LQG see \cite{thiemann_modern_2007}.} The consistency conditions reduce to three operations with respect to the edges of a given graph. (i) Deletion of edges, (ii)  invariance under inversion of edges and (iii) composition of edges. 
These consistency conditions will further restrict the form of the measure function $
mu(gg^\dagger)$. 

(i) Deletion of edges attributes to functions which are cylindrical with respect to two graphs $\gamma\subset\gamma'$, where there are edges of the larger graph that are not an argument of the function. We get consistency when these additional edges do not contribute to $\mu(F_{\gamma'})$. With $\mu$ being a product measure that does not relate edges, this essentially requires the measure on a single edge to be a probability measure. This means
\begin{equation}\label{eq:cylcon_deletion}
    \int_\SLtwoC\!\grad{g_{e\mathrm{H}}}\mu(g_e,\bar{g}_e)=1.
\end{equation}
Note that this condition is equivalent to demanding the moment $\mu_j$ for $j=0$ to be $\mu_0=1$. It normalises the reality condition implementing measure $\grad{g_{\mathrm{H}}}\mu(g,\bar{g})$. In turn, having a normalised measure here is the same requirement as taming the non-compactness of $\SLtwoC$ in order to obtain finite values from the inner product of holomorphic functions. So we see here, that there is no additional condition for the non-compactness, but it is already covered by being able to establish a cylindrically consistent measure. 

(ii) If a graph $\gamma$ is obtained from $\gamma'$ via the inversion of edges, the condition is, that $F_\gamma$ and $F_{\gamma'}$ give the same result with respect to consistent measure. This again reduces to a condition on the measure for the individual edges. While the Haar-measure $\grad{g_{e\mathrm{H}}}$ is invariant under inversion, the measure function is generally not. The inversion condition therefore gives another condition on the measure function. Hence it has to satisfy
\begin{equation}\label{eq:cylcon_inversion_0}
    \mu(g,\bar{g})=\int_\SUtwo\!\grad{h}\mu(hgg^\dagger h^{-1})=\int_\SUtwo\!\grad{h}\mu(hg^{-1}(g^{-1})^\dagger h^{-1})=\mu(g^{-1},\bar{g}^{-1}).
\end{equation}
At first we want to show that it is sufficient to have a function $\mu(gg^\dagger)$, which , before averaging, satisfies the condition $\mu(gg^\dagger)=\mu(g^{-1}(g^{-1})^\dagger)$. We look at 
\begin{equation}
    \begin{aligned}
        \int_\SUtwo\!\grad{h}\mu(hgg^\dagger h^{-1})&=\int_\SUtwo\!\grad{h}\mu(hgh^{-1}(hgh^{-1})^\dagger)\\
        &=\int_\SUtwo\!\grad{h}\mu\rb{(hgh^{-1})^{-1}((hgh^{-1})^{-1})^\dagger}\\
        &=\int_\SUtwo\!\grad{h}\mu\rb{hg^{-1}(g^{-1})^\dagger h^{-1}},
    \end{aligned}
\end{equation}
where we used unitarity of $h\in\SUtwo$ in the first line to bring the averaging into a more practical form and used $\mu(g,\bar{g})=\mu(g^{-1},\bar{g}^{-1})$ from the first to the second line. Therefore group averaging is not necessary for satisfying the inversion condition. A restriction on the function we average in order to satisfy the first reality condition is indeed enough.
Disregarding the group averaging, functions of the form 
\begin{equation}\label{eq:cylcon_inversion_1}
    \mu(g,\bar{g})=\mu\rb{gg^\dagger+g^{-1}(g^{-1})^\dagger}
\end{equation}
satisfy this condition, as $gg^\dagger+g^{-1}(g^{-1})^\dagger$ is mapped to itself under $g\mapsto g^{-1}$.

When we however take the group averaging into account and, in particular, use the symmetry $gg^\dagger\leftrightarrow g^\dagger g$, we find that 
\begin{equation}
    \int_\SUtwo\!\grad{h}\mu(hg^{-1}(g^{-1})^\dagger h^{-1})=\int_\SUtwo\!\grad{h}\mu(h(g^{-1})^\dagger g^{-1} h^{-1})=\int_\SUtwo\!\grad{h}\mu\rb{(hgg^\dagger h^{-1})^{-1}}. 
\end{equation}
Therefore condition turns into 
\begin{equation}
    \int_\SUtwo\!\grad{h}\mu\rb{(hg)(hg)^\dagger }=\int_\SUtwo\!\grad{h}\mu\rb{((hg)(hg)^\dagger)^{-1}}.
\end{equation}
As further $hg\in\SLtwoC$, it is sufficient to have a function $\mu$, satisfying 
\begin{equation}\label{eq:cylcon_inversion_2}
    \mu\rb{gg^\dagger}=\mu\rb{(gg^\dagger)^{-1}}.
\end{equation}

We do not want to go further into details about the classification of such functions that are invariant under edge inversion. Instead we give an intuitive example, which is valid for both types of functions, \eref{eq:cylcon_inversion_1} and \eref{eq:cylcon_inversion_2}, presented above. Because\footnote{This is easy to see, as we have to invert a $2\times 2$-matrix with unit determinant, i.\,e. we only change the order of diagonal elements. This does not change the trace.} 
\begin{equation}
    \tr\rb{gg^\dagger}=\tr\rb{(gg^\dagger)^{-1}}=\tr\rb{g^{-1}(g^{-1})^\dagger}
\end{equation}
and therefore 
\begin{equation}
    \tr\rb{gg^\dagger+g^{-1}(g^{-1})^\dagger}=2\tr\rb{gg^\dagger},
\end{equation}
every measure function 
\begin{equation}
    \mu(g,\bar{g})=\mu\rb{\tr(gg^\dagger)}
\end{equation}
satisfies the inversion condition. Note that the group averaging is trivial, since the $\SUtwo$ cancel via the cyclicity of the trace. 

(iii) The final consistency condition describes graphs where we replace two, $e_2$ and  $e_1$ such that $h_{e_2}h_{e_1}=h_{e_2\circ e_1}$, by a single edge, namely their concatenation $e_2\circ e_1$. The problem here is again that the measure function is not invariant under translations, but only the Haar-measure is. For simplicity, we just consider a function that is cylindrical with respect to $\gamma=\{h_{e_1},h_{e_2}\}$ and $\gamma'=\{h_{e_2\circ e_1}\}$. 

We start with the expansion of the analytic cylindrical function in terms of holomorphic irreps. This yields
\begin{equation}
   \begin{aligned}
        F_{h_{e_1\circ e_1}}(A)&=\sum_{j,m,n} \T{c}{_{jm}^n} \IR{j}{h_{e_2\circ e_1}(A)}{^m_n}\\
        &= \sum_{j,m,n} \T{c}{_{jm}^n} \IR{j}{h_{e_2}(A)h_{e_1}(A)}{^m_n}\\
        &= \sum_{j,m,n} \T{c}{_{jm}^n} \IR{j}{h_{e_2}(A)}{^m_k}\IR{j}{h_{e_1}(A)}{^k_n}\\
        &= F_{h_{e_1}, h_{e_2}}(A).
   \end{aligned}
\end{equation}
We now compute $\mu(F_{h_{e_1\circ e_1}})$ and $\mu(F_{h_{e_1}, h_{e_2}})$ individually. Starting with $F_{h_{e_1\circ e_1}}$, we find 
\begin{equation}
    \mu(F_{h_{e_1\circ e_1}})=\int_{\SLtwoC}\!\grad{g_{e_2\circ e_1 \mathrm{H}}} \mu\rb{g_{e_2\circ e_1},\overline{g_{e_2\circ e_1}}}  \sum_{j,m,n} \T{c}{_{jm}^n} \IR{j}{g_{e_2\circ e_1}}{^m_n}.
\end{equation}
The integration can be understood as the inner product of $\IR{j}{g_{e_2\circ e_1}}{^m_n}$  and the irreducible representation $\IR{0}{g_{e_2\circ e_1}}{^0_0}$, which results in  
\begin{equation}
    \mu(F_{h_{e_1\circ e_1}})=\sum_{j,m,n} \T{c}{_{jm}^n} \frac{\mu_j}{2j+1}\delta_{j,0}\delta^{m,0}\delta_{n,0}=\T{c}{_{00}^0}.
\end{equation}
Here we assumed that $\mu_0=1$ by the consistency condition with respect to deletion of edges. On the other hand we find 
\begin{equation}
\begin{aligned}[t]
    \mu(F_{h_{e_1}, h_{e_2}})=\int_{\SLtwoC^2}&\!\grad{g_{e_2\mathrm{H}}}\mu\rb{g_{e_2},\overline{g_{e_2}}}\grad{g_{e_1\mathrm{H}}}\mu\rb{g_{e_1},\overline{g_{e_1}}}\\ &
    \times\sum_{j,m,n} \T{c}{_{jm}^n} \IR{j}{h_{e_2}(A)}{^m_k}\IR{j}{h_{e_1}(A)}{^k_n}.
    \end{aligned}
\end{equation}
Performing the $g_{e_2}$ integration as above yields 
\begin{equation}
    \mu(F_{h_{e_1}, h_{e_2}})= \int_{\SLtwoC}\!\grad{g_{e_1\mathrm{H}}}\mu\rb{g_{e_1},\overline{g_{e_1}}} \sum_{j,m,n} \T{c}{_{jm}^n} \frac{\mu_j}{2j+1} \delta_{j,0}\delta^{m,0}\delta_{k,0} \IR{j}{h_{e_1}(A)}{^k_n}.
\end{equation}
With $j=0$, $\IR{j}{h_{e_1}(A)}{^k_n}$ is set to the trivial representation and $k$ only takes the value $0$. Hence the remaining $g_{e_1}$-integration again results in $\mu_0=1$. Therefore we end up with 
\begin{equation}
    \mu(F_{h_{e_1}, h_{e_2}})=\T{c}{_{00}^0}=\mu(F_{h_{e_1\circ e_1}}).
\end{equation}
The consistency condition for composition of edges is indeed already satisfied. There are two aspects that contributed to this result. First, only because we have a measure with respect to which the holomorphic irreps are orthogonal, we are able to obtain $\mu(F_{h_{e_1}, h_{e_2}})=\mu(F_{h_{e_1\circ e_1}})$. As this orthogonality is induced by the implementation of RCI, in a sense only the reality condition makes it possible to obtain a cylindrically consistent measure. The second aspect concerns the occurring moments and the consistency in regard to the deletion of edges. Without this, we would have $\mu(F_{h_{e_1\circ e_1}})\sim\mu_0$, while $\mu(F_{h_{e_1}, h_{e_2}})\sim{\mu_0}^2$. Therefore we need $\mu_0=1$ in order to obtain consistent composition of edges. 

In summary, the first reality condition and the consistency conditions for edge deletion and edge inversion define a class of  measures
\begin{equation}
    \grad{\mu\rb{g,\bar{g}}}=\grad{g_\mathrm{H}}\mu(g,\bar{g})
\end{equation}
on the single edge holomorphic Hilbert space, from which we, similar to real LQG, can construct a consistent measures $\grad\mu_0$ on the holomorphic Hilbert space of generalised $\SLtwoC$ connections
\begin{equation}
    \curlyH_0=\curlyL^2\rb{\bar{\curlyA},\grad\mu_0}\cap\mathfrak{H}\rb{\bar{\curlyA}}.
\end{equation}

Note that the conditions \eref{eq:cylcon_deletion} for edge deletion and non-compactness and \eref{eq:cylcon_inversion_0} for edge inversion still have to be both satisfied independently. We will not analyse their general compatibility here. Instead we look at the previous example regarding $\tr\rb{gg^\dagger}$. Even with this, every normalisable function $\mu\rb{\tr\rb{gg^\dagger}}$ results in an appropriate measure. 

\subsection{$\SLtwoC$ spin networks and gauge invariance}\label{sec:spin_networks_gauge_invariance}
Now we want to generalise the notion of spin network functions to $\SLtwoC$. We do not want to go into the details of the orthogonal decomposition of $\curlyH_0$ and refer to the standard literature, e.\,g. \cite{ashtekar_background_2004,giesel_classical_2013}. 

The important point is that from the holomorphic and normalised irreps we can construct spin network functions $T_{\gamma jmn}(A)$, based on a graph $\gamma$ and the $\SLtwoC$ holonomies along its edges. Namely,
\begin{equation}\label{eq:spin_network_basis}
    T_{\gamma jmn}(A)=\prod_{e\in E(\gamma)}  \bIR{j_e}{h_e(A)}{_{m_e n_e}},
\end{equation}
with $j=\{j_e\}$, $m=\{m_e\}$ and $n=\{n_e\}$, the collections of representation labels. Because of the orthogonality of the holomorphic irreps, they provide an orthonormal basis of the graph Hilbert space $\curlyH_\gamma$ which consists of holomorphic functions of $\SLtwoC^{|E(\gamma)|}$. The inner product is determined by the product measures for the individual edges as in \eref{eq:cylcon_measure}.

In this sense the spin network functions constructed from the basis \eref{eq:spin_network_basis}, behave exactly like the spin network functions of real LQG, but with the difference that we now have spin network functions with respect to $\SLtwoC$ holonomies instead of $\SUtwo$ holonomies. 

Finally we want to consider gauge invariance of these $\SLtwoC$ spin network states. 
We recall the Clebsch-Gordan decomposition of the tensor product of  irreducible representations of $\SUtwo$, i.\,e.  
\begin{equation}\label{eq:CG_decomp}
    \Pi_{j_1}\otimes\Pi_{j_2}=\bigoplus_{j=|j_1-j_2|}^{j_1+j_2} \Pi_j, 
\end{equation}
in terms of the direct sum of irreducible representations. For the representation matrices $\IR{j}{h}{^m_n}$, where $h\in\SUtwo$ this decomposition takes the well known form
\begin{equation}
    \IR{j_1}{h}{^{m_1}_{n_1}}\IR{j_2}{h}{^{m_2}_{n_2}}=\sum_{j_3=|j_1-j_2|}^{j_1+j_2}\sum_{m_3,n_3=-j_3}^{j_3} \CGC{j_1j_2j_3}{^{m_1 m_2}_{m_3}}\CGC{j_1j_2j_3}{_{n_1 n_2}^{n_3}} \IR{j_3}{h}{^{m_3}_{n_3}}.
\end{equation}
Here $\CGC{j_1j_2j_3}{_{n_1 n_2}^{n_3}}$ are the Clebsch-Gordan coefficients. As usual, they are real and only non-vanishing when the Clebsch-Gordan conditions are satisfied. 

The crucial point is that everything considered here are irreducible representations of $\SUtwo$. Therefore Weyl's unitary trick allows us to lift the $\SUtwo$ irreps to holomorphic irreducible representations of $\SLtwoC$. Thus the Clebsch-Gordan decomposition \eref{eq:CG_decomp} holds for the holomorphic irreps of $\SLtwoC$ as well. Further the decomposition of products of representation matrices holds also for $g\in\SLtwoC$, i.\,e.
\begin{equation}
    \IR{j_1}{g}{^{m_1}_{n_1}}\IR{j_2}{g}{^{m_2}_{n_2}}=\sum_{j_3=|j_1-j_2|}^{j_1+j_2}\sum_{m_3,n_3=-j_3}^{j_3} \CGC{j_1j_2j_3}{^{m_1 m_2}_{m_3}}\CGC{j_1j_2j_3}{_{n_1 n_2}^{n_3}} \IR{j_3}{g}{^{m_3}_{n_3}}.
\end{equation}
As a result, the holomorphic irreducible representations of $\SLtwoC$ feature the same recoupling theory as the $\SUtwo$ irreducible representations.  

In real LQG the Gau{\ss} constraint is solved by the use of spin network function that carry invariant tensors at their vertices, such that the spins of incoming and outgoing edges at all individual vertices of the graph couple to $0$. For a detailed description, which works exactly the same for the Gau{\ss} constraint in \eref{eq:classical_constraints} here, we refer to \cite{ashtekar_isolated_2004,thiemann_modern_2007,giesel_classical_2013}. Because now $\SUtwo$ and holomorphic $\SLtwoC$ irreps share the same recoupling theory, we can use exactly the same method in order to make $\SLtwoC$ spin network functions gauge invariant. 

We want to comment on the inner product \eref{eq:inner_product_one_edge_HS} on the single edge Hilbert space. With an RCI implementing measure \eref{eq:RCI_measure}, we realised that the inner product is not invariant under $\SLtwoC$ transformation, but only under its rotational subgroup $\SUtwo$. At first glance this seems to be problematic and might even suggest a restriction of the theory to $\SUtwo$. With what we now understand about the invariance of the spin network function under $\SLtwoC$ gauge transformations, there is no restriction to this subgroup necessary, when considering gauge invariance on the Hilbert space $\curlyH_0.$

\section{Discussion and outlook}
\label{sec:discussion}
In the present article we describe the first 
%essential 
steps towards a formulation of selfdual loop quantum gravity. This includes the construction of the kinematical Hilbert space with a cylindrically consistent inner product that implements the selfadjointness of fluxes, i.\,e., the first reality condition, which classically corresponds to the reality of the spatial metric. The implementation of the second reality condition, which is supposed to keep the spatial metric real, will be considered elsewhere \cite{sahlmann_revisiting_RCII}. 

Starting with just a single copy of $\SLtwoC$, corresponding to a single edge and the corresponding holonomy, it was possible to obtain an inner product that ensures selfduality of flux operators. Even though we found a general solution for our ansatz for the inner product of the single edge Hilbert space, this might, in fact, not be the most general ansatz. In particular, our measure function is independent of any information, like length, about the edges, which we are actually integrating over. Inclusion of such edge dependence in real LQG allows for fluctuations of the electric field, but prevents the flux operators from being selfadjoint and causes other problems. In this holomorphic setup, however, the selfadjointness might not pose a problem, as here all action of the fluxes can be transferred to the measure function. 

Another type of measure, which we did not explore here, is used in \cite{thiemann_account_1995,thiemann_reality_1996}, where the inner product on the Hilbert space of complex connections is obtained from a Wick rotation, facilitated by a complexifier. It is therefore inherently linked to the inner product of the underlying real theory. The obstruction of this approach, however, is the existence of such a complexifier, cf. also \cite{thiemann_modern_2007,varadarajan_euclidean_2019}. For us it would nevertheless be interesting to analyse such kind of measure, that results from the Wick rotation, and whether or not there is a relation to our approach, without the an underlying real theory. Further, the measure used in \cite{wilson-ewing_loop_2015} can be brought to a somewhat similar form. However it is not a priori clear whether or not such complexifier-type measures, as in \cite{wilson-ewing_loop_2015}, allow for a finite inner product, without additional structure.  

Restricting to the Hilbert space formulation and the implementation of flux selfadjointness on the single edge Hilbert space of holomorphic $\SLtwoC$ functions, we were able to transfer known structures from real LQG to this selfdual formulation, even without the second reality condition: 
\begin{itemize}
\item With an inner product, that implements flux selfadjointness, the holomorphic irreducible representations of $\SLtwoC$ serve as a basis for the single edge Hilbert space.
\item From the holomorphic irreps we can construct an $\SLtwoC$ spin network basis, which has exactly the same recoupling theory as $\SUtwo$ spin networks. This allows to use the known method of making spin networks gauge invariant, namely by attaching to each vertex an intertwiner projecting onto the trivial representation. 
\item With a product ansatz, where each edge of a spin network is associated with a copy of the RCI implementing measure, a new class of cylindrically consistent  measures on the Hilbert space of generalised $\SLtwoC$ connections is established. 
\end{itemize}

Everything presented in this work was done with to a first reality condition, that is more restrictive than the classical formulation requires. Considering an electric field, where the imaginary part is caused only by a gauge rotation, cf. \cite{ashtekar_gravitational_2021,sahlmann_revisiting_classical_theory}, it might be possible to relax the requirement of selfdjoint fluxes to \emph{selfadjoint} up to gauge. This might open different possibilities for finding an inner product. However, more work in that direction is necessary, in order to give a clear statement. 

The next step towards a version of loop quantum gravity, based on selfdual variables and therefore $\SLtwoC$, is the implementation of the second reality condition. Classically this is the reality of the extrinsic curvature. With the regularisation methods used in LQG, however, we do not have direct access to this as an operator and therefore cannot implement it with a similar procedure as compared to the fluxes for the first reality condition. In \cite{sahlmann_revisiting_RCII} we will consider it as an adjointness relation for holonomies (restricting the moments of the measure), as well as a restriction on the wave functions.

\ack
R.S. and H.S. benefitted from discussions with Wojciech Kamiński, Jerzy Lewandowski, Thomas Thiemann, Madhavan Varadarajan and Wolfgang Wieland. H.S. acknowledges the contribution of the COST Action CA18108. R.S. thanks the Evangelisches Studienwerk Villigst for financial support.

%\bibliographystyle{unsrt.bst}
%\bibliographystyle{iopart-num}
%\bibliography{references.bib}

\end{document}

%% file: lib.tex
\newcommand{\CGC}[2]{\tensor{{c^{(#1)}}}{#2}}
\newcommand{\IR}[3]{\Pi_{#1}\tensor{(#2)}{#3}}

\newcommand{\bIR}[3]{b_{#1}\tensor{(#2)}{#3}}

\renewcommand{\Im}{\mathrm{{Im}}}

\newcommand{\SLtwoC}{{\mathrm{SL}(2,\mathbb{C})}}	
\newcommand{\sltwoC}{\mathfrak{sl}(2,\mathbb{C})}

\newcommand{\LGC}{\mathrm{SO}(3,1;\mathbb{C})}

\newcommand{\RR}{\mathbb{R}}
\newcommand{\CC}{\mathbb{C}}

\newcommand{\T}[2]{\tensor{#1}{#2}}

\newcommand{\lc}[1]{\tensor{\epsilon}{#1}}

\newcommand{\tE}{\tilde{E}}

\newcommand{\grad}[1]{\mathrm{d}{#1}}

\newcommand{\curlyA}{\mathcal{A}}

\newcommand{\curlyH}{\mathcal{H}}
\newcommand{\curlyL}{\mathcal{L}}

\newcommand{\curlyP}{\mathcal{P}}

\newcommand{\rb}[1]{\left({#1}\right)}

\newcommand{\curlybrackets}[1]{\left\{{#1}\right\}}

\newcommand{\SUtwo}{{\mathrm{SU}(2)}}			% SU(2)
\newcommand{\Uone}{\mathrm{U}(1)}		    % U(1)
\newcommand{\sutwo}{\mathfrak{su}(2)}			% su(2)
 
\newcommand{\commutator}[2]{\left[{#1},{#2}\right]}			% commutator
\renewcommand{\PB}[2]{\left\{#1,#2\right\}}	% Poisson bracket
\newcommand{\innerproduct}[2]{\langle{#1,#2}\rangle}	% inner product